\documentclass[referee]{aa} % for a referee version
\usepackage{subfigure}
\usepackage{graphicx}
\usepackage{txfonts}
\begin{document}
\newcommand{\rmn}[1] {{\rm #1}}
\newcommand{\Zsolar}{\mbox{${\; {\rm {\sun}}}$}}
\newcommand{\ha}{\hbox{H$\alpha$}}
\newcommand{\oii}{\hbox{[O\,{\sc ii}]}}
\newcommand{\neiii}{\hbox{[Ne\,{\sc iii}]}}
\newcommand{\siii}{\hbox{[S\,{\sc iii}]}}
\newcommand{\ariii}{\hbox{[Ar\,{\sc iii}]}}
\newcommand{\nii}{\hbox{[N\,{\sc ii}]}}
\newcommand{\sii}{\hbox{[S\,{\sc ii}]}}
\newcommand{\oiii}{\hbox{[O\,{\sc iii}]}}
\newcommand{\hb}{\hbox{H$\beta$}}
\newcommand{\hg}{\hbox{H$\gamma$}}
\newcommand{\hd}{\hbox{H$\delta$}}
\newcommand{\hi}{\hbox{H\,{\sc i}}}
\newcommand{\hii}{\hbox{H\,{\sc ii}}}
\newcommand{\heii}{\hbox{He\,{\sc ii}}}
\newcommand{\etal}{\hbox{et\thinspace al.\ }}
\newcommand{\oiiha}{\hbox{[O\,{\sc ii}]/H$\alpha$}}
\newcommand{\fha}{\hbox{$F_{{\rm H}\alpha}$}}
\newcommand{\fhb}{\hbox{$F_{{\rm H}\beta}$}}
\newcommand{\foii}{\hbox{$F_{\rm [O\,{\sc II}]}$}}
\newcommand{\micron}{\hbox{$\mu$m}}
\newcommand{\zoh}{\hbox{$12\,+\,{\rm log(O/H)}$}}
\newcommand{\teoii} {\hbox{$T_e{\rm (O\,{\sc II})}$}}
\newcommand{\tenii} {\hbox{$T_e{\rm (N\,{\sc II})}$}}
\newcommand{\teoiii}{\hbox{$T_e{\rm (O\,{\sc III})}$}}
   \title{ Artificial Neural Network  for  search for  metal poor galaxies}

  % \subtitle{I. Overviewing the $\kappa$-mechanism}

   \author{Fei,Shi
             \inst{1}
           \and
          Yu-Yan, Liu \inst{1}
           \and
           Xu, Kong \inst{2,3}
           \and
            Yang, Chen
           \inst{2}
           }

   \institute{North China Institute of Aerospace Engineering, Langfang,
Hebei, 065000, China.\\ \email{fshi@bao.ac.cn}
         \and
             Center of Astrophysics, University of Science and
Technology of China, Jinzhai Road 96, Hefei 230026, China.\\
\email{xkong@ustc.edu.cn}
         \and
             Key Laboratory for Research in Galaxies and Cosmology,
USTC, CAS, China.
            }

   \date{Received XXXX, 2013; accepted XXXX, 2013}

% \abstract{}{}{}{}{}
% 5 {} token are mandatory

  \abstract
   {}
   { In order to find a fast and reliable method for selecting metal poor galaxies (MPGs),
   especially in large surveys and huge database,
   an Artificial Neural Network (ANN) method is applied
   to a sample of star-forming galaxies from the Sloan Digital Sky Survey (SDSS) data release 9 (DR9) provided by
the Max Planck Institute  and the Johns Hopkins University (MPA/JHU)\footnote{http://www.sdss3.org/dr9/spectro/spectroaccess.php}.
   }
   {A two-step approach is adopted:(i) The ANN network must  be ¡°trained¡± with
a subset of objects that are known to be either MPGs or  MRGs(Metal Rich galaxies),
 treating the strong emission line flux measurements as input feature vectors in an
n-dimensional space, where n is the number of strong emission line flux ratios.
(ii)  After the network is trained on a sample of star-forming galaxies,
remaining galaxies are classified in the automatic test analysis as either MPGs or MRGs.
We consider several random divisions of the data into
training and testing sets: for instance, for our sample, a total of 70  per cent of
the data are involved in training the algorithm, 15 percent are involved in validating the algorithm
and the remaining 15 percent are used for blind testing of the resulting classifier.
   }
  % results heading (mandatory)LGM_TOT_P2P5
   {For target selection, we have achieved an acquisition rate for MPGs
of 96 percent and 92 percent for an MPGs threshold of \zoh=8.00 and \zoh=8.39 ,
respectively.
{\bf Running the code takes minutes in most cases under the Matlab 2013a software environment.}
The code in the paper is available at the web\footnote{http://fshi5388.blog.163.com}.
The ANN method can easily be extended to any MPGs target selection task
when the physical property of the target can be expressed as a quantitative variable.
}
   {}


   \maketitle
%
%________________________________________________________________

\section{Introduction}

Extremely metal poor local galaxies are key in an understanding of
star formation and enrichment in a nearly
pristine interstellar medium (ISM). Metal poor galaxies
provide important constraints on the pre-enrichment
of the ISM by previous episodes of star formation, e.g by Population
III stars (Thuan et al. 2005). Metal poor galaxies are also the best objects for
the determination of the primordial He abundance and for constraining cosmological models (e.g.
Izotov \& Thuan 2004; Izotov et al. 2007). Metal poor galaxies
are possibly the closest examples we can find of elementary primordial units from which
galaxies formed.

Unfortunately, MPGs are rare. The review by Kunth \& {\" O}stlin (2000) cites only
31 targets with metallicity below one tenth the solar value.
The first extragalactic objects with very low metal abundance
were discovered by Searle \& Sargent (1972) who reported on the
properties of two intriguing galaxies, IZw18 and IIZw40.
Their extreme metal under-abundance, more than 10 times less than
solar, and even more extreme than that of H{\sc ii} regions found
in the outskirts of spiral galaxies, indicates
that these objects could genuinely be young galaxies
 in the process of formation (Kunth \& {\" O}stlin 2000).
This discovery leads to extensive systematic searches for more objects
with low metallicity (see Kunth \& {\" O}stlin 2000 and references
therein) to understand the  properties of their massive stars
(formation and evolution, appearance of WR stars), the evolution
of the dynamics of the gas in the gravitational potential of the
parent galaxy as a superbubble evolves, the triggering mechanism
that ignites their bursts of star formation, and the chemical enrichment
of the interstellar medium after the fresh products are well mixed.

The number of MPGs has significantly increased since the work by Kunth \& {\" O}stlin (2000),
 but still the number of known MPGs is small (Morales-Luis et al.2011).
 The thorough bibliographic compilation described in ¡ì 4 shows only 421 MPGs
 with metallicity below two tenth the solar value ( $\zoh<8.0$).
There are several reasons that prohibit the identification of more MPGs.
One reason  is that the MPGs are usually dwarf galaxies which are dim and hard to observe.
Another reason is that the methods that determine the  metallicity of a galaxy are highly uncertain
and there is a large discrepancy between different methods (Shi et al.2005, 2006, 2010).

The metallicity is a key parameter in the search for MPGs. Oxygen is an
important element that is easily and reliably determined since
 the most important ionization stages can  be observed.
The oxygen abundance from the measurement of electron temperature
from  \oiii$\lambda\lambda$4959,5007/\oiii$\lambda$4363 is
one of the most reliable methods,  so called the $T_e$ method.
But \oiii$\lambda4363$ is usually weak in low metallicity galaxies,
and there are often large errors when measuring this line. In high
metallicity galaxies, \oiii$\lambda4363$ is hardly even observable.

Instead of the $T_e$ method, strong line methods,  such as the $R_{23}$
 \footnote{$R_{23}$=(\oii$\lambda$3727+\oiii$\lambda\lambda$4959,5007)/\hb},
$P$ \footnote{$P$=\oiii$\lambda\lambda$4959,5007/(\oii$\lambda$3727+\oiii$\lambda\lambda$4959,5007)},
$N2$ \footnote{$N2$=log(\nii$\lambda$6583/\ha)},
$Ne3O2$ \footnote{$Ne3O2$=log(\neiii$\lambda3869$/\oii$\lambda3727$)}, or
$O3N2$ \footnote{$O3N2$=log((\oiii$\lambda$5007/\hb)/(\nii$\lambda$6583/\ha))} methods, are widely used
 (Pagel \etal 1979; Kobulnicky, Kennicutt, \& Pizagno 1999; Pilyugin \etal 2001; Charlot
\& Longhetti 2001; Denicol{\'o} \etal 2002; Pettini \& Pagel 2004;
Tremonti \etal 2004; Liang \etal 2006). The $R_{23}$ and  $P$ methods suffer the
double-valued problem, requiring some assumption or rough  a priori
knowledge of a galaxy's metallicity in order to locate it on the
appropriate branch of the relation. The $N2$- and  $O3N2$ methods
are monotonic, but the reasons for this are not purely  physical. It is
partly due to  the N/O ratio increasing on average with the increase in
metallicity (Stasi{\'n}ska 2006; Shi \etal 2006). Besides, calibrations
of the $O3N2$ and $N2$ indices might be improper for interpreting the integrated
spectra of galaxies because $\nii\lambda6583$  and \ha\ may arise not only in bona
fide $\hii$ regions, but also in a diffuse ionized medium.
Stasi{\'n}ska (2006)   proposed $Ar3O3$ \footnote{$Ar3O3$=\ariii$\lambda7135$/\oiii$\lambda5007$}
 and $S3O3$ \footnote{$S3O3$=\siii$\lambda9069$/\oiii$\lambda5007$} as new abundance
indicators, which have the advantage of being unaffected by the effects of
chemical evolution. The advantages are superior to previous $N2$  and $O3N2$
methods.

In short,  the method using a single flux ratio is questionable.
The ideal metallicity indicator should use all the strong emission lines.
In order to use a method that capitalizes on strong emission lines to identify MPGs,
we have employed an
automatic Artificial Neural Network(ANN)  search for metal poor galaxies
in the ninth Sloan Digital Sky Survey data release (SDSS/DR9),
by combining all strong emission line flux ratio measurements including
 $Ne3O2$,\oiii$\lambda\lambda$4959,5007/\oiii$\lambda$4363, \oii/\hb, \oiii/\hb, \ha/\hb, $N2$ and \sii,
 provided by MPA.  An ANN approach has already successfully been applied to sort out different types of astronomical
spectra, from supernovae (Karpenka et al.2013) to quasars  (Y{\`e}che et al. 2010).

This paper is organized as follows. In Sect. 2, we describe
the data set used for training and testing in our analysis, and we
present a detailed account of our methodology in Sect. 3. We test
the performance of our approach in Sec. 4 by applying to the data
and present our results. Finally, we conclude in In Sect. 5.
Throughout the paper, we adopt cosmological parameters of the $\Omega_M$ = 0.27
and $\Omega_\Lambda$ = 0.73.

%__________________________________________________________________

\section{Data sample}

\begin{figure*}
   \centering
   \includegraphics[scale=0.9,angle=0]{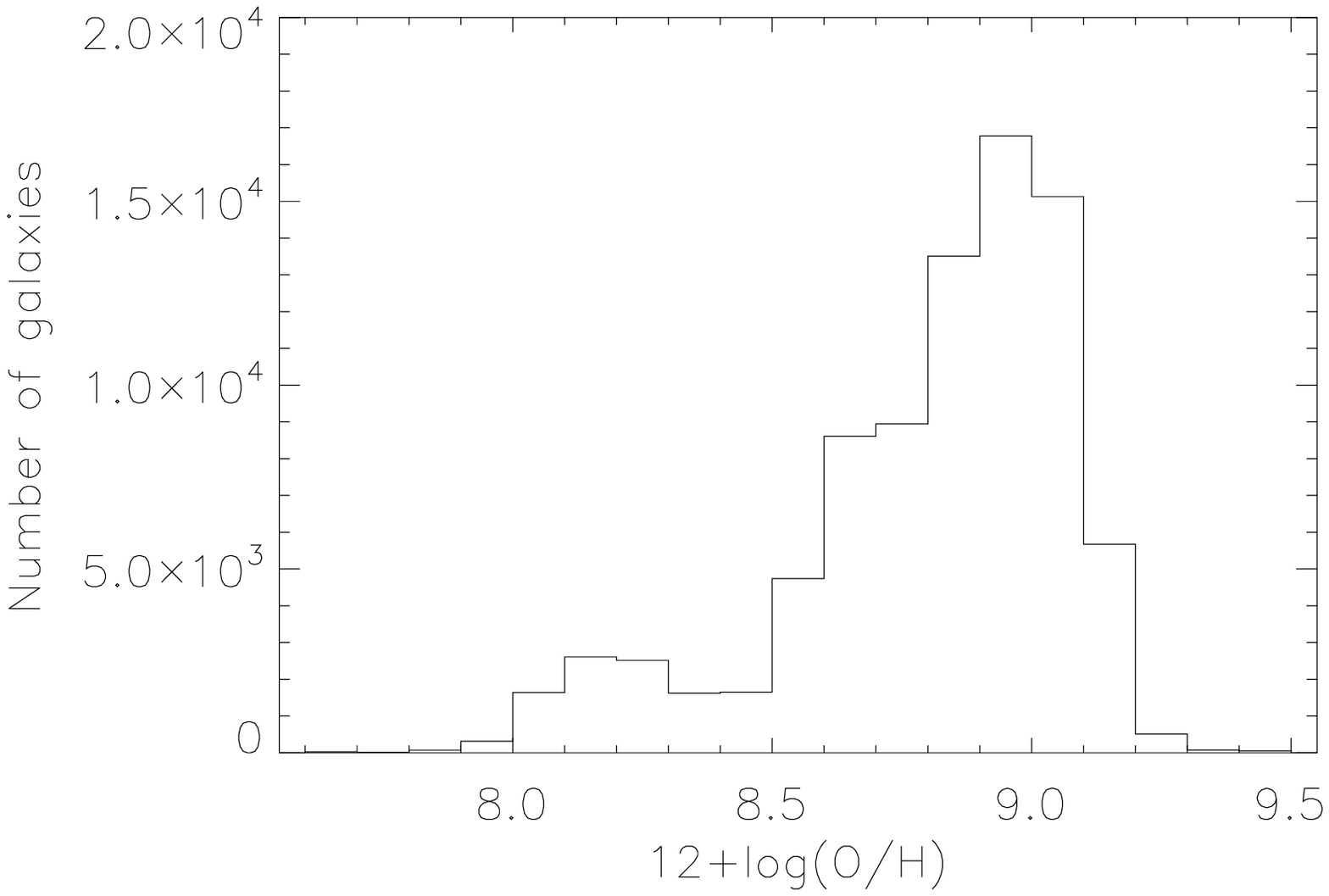}
   \caption{The distribution of \zoh for our data sample. }
              \label{Fig1}%
    \end{figure*}

   In order to use ANN, we must first construct a sample of sources with good emission line detections.
    All the objects in the sample must have evident and reliable flux ratio measurements,
    such as  $Ne3O2$,  \oiii$\lambda\lambda$4959,5007/\oiii$\lambda$4363, \oii/\hb, \oiii/\hb, \ha/\hb, \nii/\ha and \sii.
  For this purpose, we use the catalog of star-forming galaxies in SDSS
   DR 9 provided by MPA which made use of
the spectral diagnostic diagrams from Kauffmann et al. (2003)
to classify galaxies as star-forming galaxies, active galactic nuclei (AGN), or unclassified.
 In total, 84,465 star-forming galaxies are adopted in our sample.
All the galaxies in the sample have reliable spectral observations
with  reasonable values of strong line ratio and oxygen abundance.
 The redshifts of  the galaxies in the sample are in the range of   0.02 $< z <$ 0.3.
The oxygen abundance in the sample spans a wide range from 7.1 $<\zoh<$9.5.
The distribution of the oxygen abundance in each bin is plotted in Fig.\ref{Fig1}.
 It is clearly evident that MPGs are rare compared to metal rich galaxies (MRGs).
 There are only 8671 galaxies with  $\zoh<8.39$($\sim 10$ percent of the star-forming galaxies sample),
and among them only 421 galaxies with $\zoh<8.0$($\sim 0.5$ percent of the star-forming galaxies sample).

\section{ Artificial Neural Network Approach }

\begin{figure*}
   \centering
   \includegraphics[scale=0.9,angle=0]{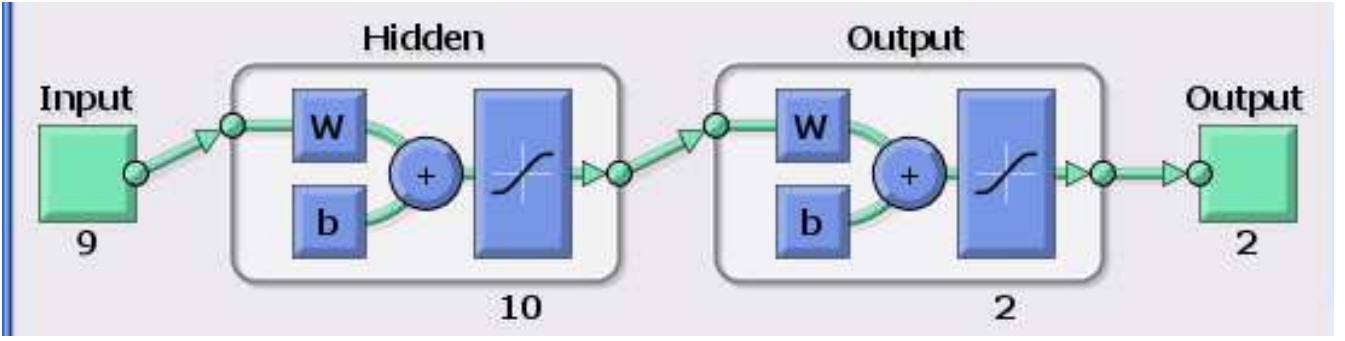}
   \caption{ Schematic representation of  a 2-layer artificial neural network used
   here with 9 input variables (redshift, $Ne3O2$,  \oiii$\lambda\lambda$4959,5007/\oiii$\lambda$4363, $N2$/\ha
,\oiii/\hb, \sii, \oii/\hb,  \ha/\hb, \ariii/\oiii) as the  input vector {\bf x},
    10 hidden neurons, and 2 output neurons (MPG or MRGs as target vectors {\bf t}). This schematic is taken from
    the MATLAB Recognizing Patterns Documentation. }
    \label{Fig2}%
    \end{figure*}

The basic building block of the ANN architecture
is a processing element called a neuron. The ANN architecture used in this
study is illustrated in Fig.\ref{Fig2}. For target selection,
we use an {\bf nprtool} package developed in the Matlab environment.
The {\bf nprtool} package leads us through solving a pattern-recognition classification problem
using a two-layer feed-forward  network with sigmoid output neurons.
The neuron is the simplest kind of node, and maps an input vector ${\bf x} \in \Re^n$ to
a scalar output $f({\bf x};{\bf w},\theta)$ via
\begin{equation}
\label{eq:perceptron}
f({\bf x};{\bf w},\theta) = \theta + \sum_{i=1}^n {w_{i} x_{i}},
\end{equation}
 where {\bf $\theta$} are `bias' and
 {\bf $\{w_{i}\}$}  are  `weights' of the i-th variables in the input vector {\bf x}
 which include {\bf n} variables.  We will focus mainly on a
2-layer feed-forward ANN, which consists of  a hidden layer and an
output layer as shown in Fig. 2.
The default number of hidden neurons is set to 10. The number of output neurons is set to 2,
which is equal to the number of elements in the target vector (MPGs or MRGs).

In order to classify a set of data using an ANN, we need to provide
a set of training data ({\bf x t}).
We build the input vector {\bf x}, that includes redshift, $Ne3O2$, \oiii$\lambda\lambda$4959,5007/\oiii$\lambda$4363, $N2$/\ha
,\oiii/\hb, \sii, \oii/\hb, \ha/\hb, and \ariii/\oiii data ( 9 input variables)
from the data-set in Section 2.
Target {\bf t} is defined as 0 or 1 to represent MPGs or MRGs.
We applied an MPGs cut to \zoh=8.0 (corresponding to 0.2Z$_{\sun}$), to enhance the selection.
Because MPGs are much rarer than MRGs, we select  1,000  MRGs galaxies for \zoh $>$8.0 randomly
  to avoid systematic errors caused by too many MRGs compared to MPGs.

Then,  the input vectors  {\bf x} and target vectors {\bf t}
are randomly divided into three sets as follows:
70 percent are the training set, which is used for computing the gradient and updating the network weights and biases.
15 percent are used to validate that the network is generalizing and to stop training before over-fitting.
The last 15 percent are used as a completely independent test of network generalization.

Once the network weights and biases are initialized, the network is ready for training.
The process of training a neural network involves tuning the values of the weights and biases
of the network to optimize network performance, as defined by the network performance function.
The default performance function for feed-forward networks is mean squared error ({\bf $\varepsilon$})
between the network outputs {\bf f} and the target outputs {\bf t}.
It is defined as follows:

\begin{equation}
 \varepsilon = \frac{1}{2} \sum_k (f_k - t_k)^2,
\end{equation}

Depending on the network architecture, there can be millions
of network weights and biases which makes network training a very
complicated and computationally challenging task.
The {\bf nprtool} uses the simplest optimization algorithm - gradient descent.
It updates the network weights and biases in the direction in which
 the performance function decreases most rapidly, the negative of the gradient.
 The gradient will become very small as the training reaches a minimum of the performance.
 The iteration of this algorithm can be written as
\begin{equation}
 x_{k+1} = x_{k} - a_k g_k,
\end{equation}
where {\bf $x_k$} is a vector of current weights and biases, {\bf $g_k$} is the current gradient,
 and {\bf $a_k$} is the learning rate.

\section{Spectral  selection of MPGs}

Once the network has been trained, it is applied to the testing dataset
to obtain the predictions for each galaxy therein being either a MPG or a MRG.
For illustration, we considered three ANN configurations that
differ in terms of the number of variables. The first one uses
all variables:  redshift, $Ne3O2$,  \oiii$\lambda\lambda$4959,5007/\oiii$\lambda$4363, $N2$/\ha
,\oiii/\hb, \sii, \oii/\hb, \ha/\hb, and \ariii/\oiii.
In the second configuration,
We study $T_e$ method, strong line methods,  such as the $R_{23}$,
$Ar3O3$ , $N2$ , $Ne3O2$, or $O3N2$
one by one to show   which strong line ratios are  most effective in the identification of MPGs.
In the third configuration,   we plot in Fig.5 the  confusion matrices
 when the MPG threshold is fixed to \zoh=8.39(corresponding to 0.5Z$_{\sun}$)
 using all 9 variables to show the changes  caused by the MPG threshold.
The confusion matrices of   configurations are plotted in Figures 3 and 5, respectively.
For the introduction to confusion matrices, please see the Matlab Recognizing Patterns  web site
\footnote[1]{http://www.mathworks.com}.

 For the first configuration (i.e., using all variables),
 we achieve an MPG acquisition rate  of $\sim$96 percent for  MPG threshold
of \zoh=8.0. It is therefore apparent that the 9-variable ANN should
be used for the purpose of selecting MPGs in any optical spectral catalog.
We also plot the {\bf Receiver Operating Characteristic}
(ROC) curves  for our analysis
procedure in Fig. \ref{roc}. The ROC curve provides a very reliable way of
sorting out the optimal algorithm in signal detection theory.
The ROC curve is a plot of the true positive rate (sensitivity) versus
 the false positive rate (1 - specificity).
A perfect test would show points in the upper-left corner,
with 100 percent accuracy.
One sees in Fig. 4 that nprtool yields very reasonable ROC curves,
 indicating that the classifiers are quite discriminative.

\begin{figure*}
   \centering
   \includegraphics[scale=0.9,angle=0]{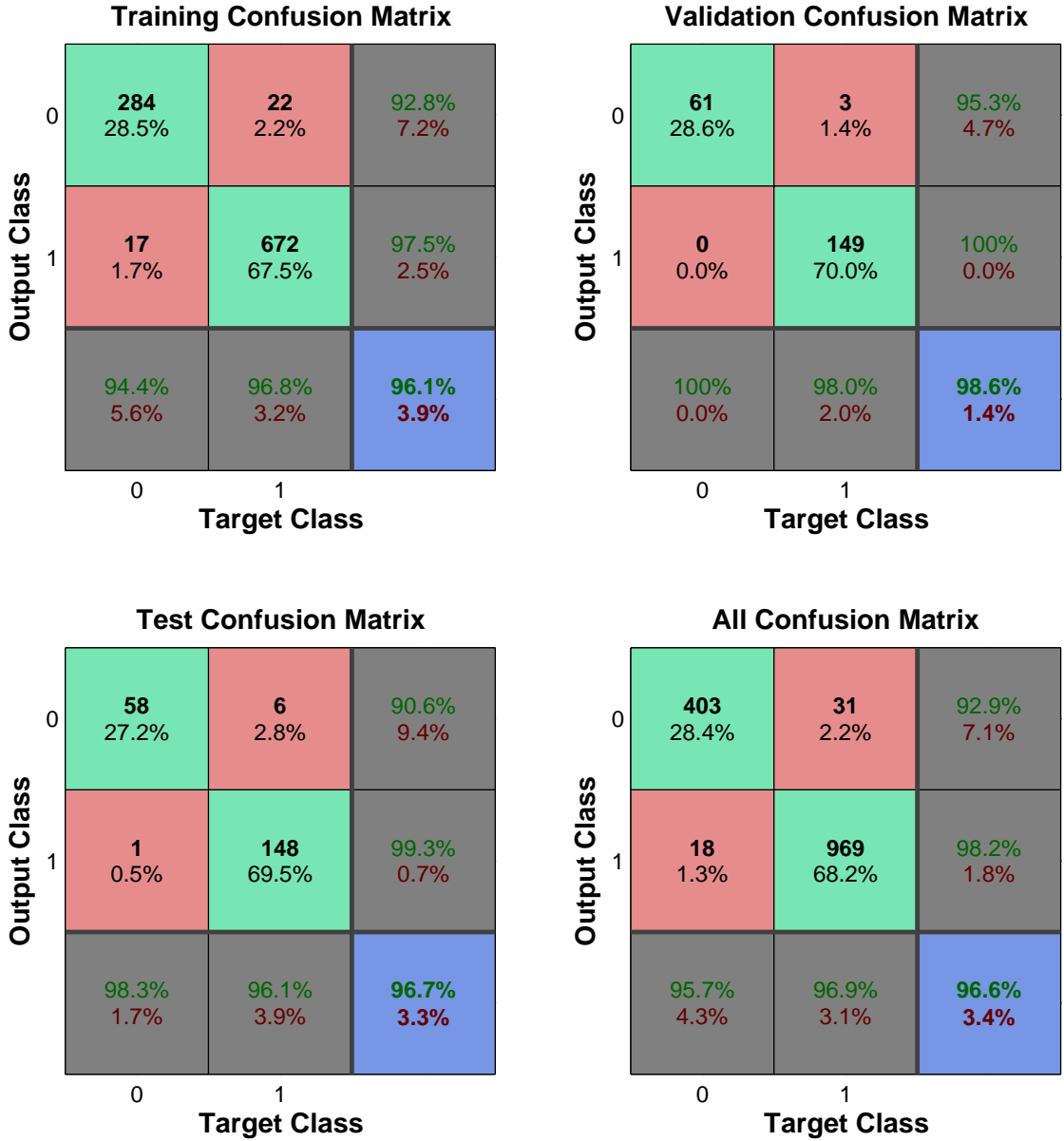}
   \caption{ the confusion matrices for training, testing, and validation,
   and the three kinds of data combined for setting the MPG threshold to \zoh=8.0 using all 9 variables.
   The diagonal cells show the number of cases that were correctly classified,
    and the off-diagonal cells show the misclassified cases.
    The blue cell in the bottom right shows the total percent of correctly classified cases (in green)
    and the total percent of misclassified cases (in red). }
    \label{Fig3}%
    \end{figure*}

\begin{figure*}
   \centering
   \includegraphics[scale=0.9,angle=0]{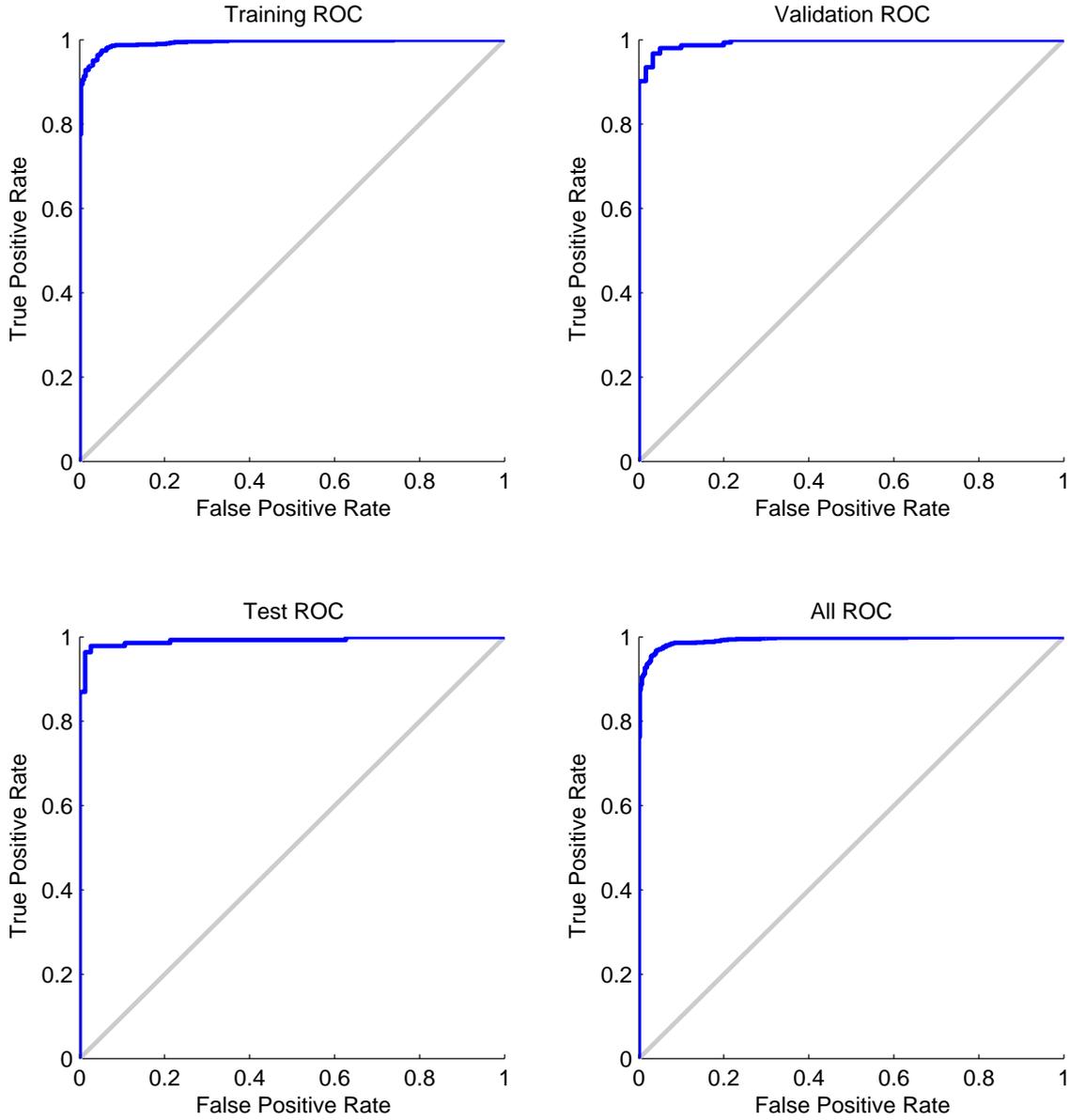}
   \caption{  The colored lines in each axis represent  Receiver Operating Characteristic (ROC) curve
   for training, testing and validation, and the three kinds of data are combined for setting the MPG threshold to \zoh=8.0.}
    \label{roc}%
    \end{figure*}

  For the sencond configuration, we identify MPGs only using the essential information for $T_e$-method
 ( redshift,  \oiii$\lambda\lambda$4959,5007/\oiii$\lambda$4363, \oiii/\hb, \sii, \oii/\hb),
  $R_{23}$-method (redshift, \oiii/\hb, \oii/\hb), $N2$-method(redshift, $N2$/\ha),
  $O3N2$-method(redshift, $N2$/\ha, \oiii/\hb ), $Ne3O2$-method(redshift, \neiii$\lambda3869$/\oii$\lambda3727$ )
 $Ar3O3$- method(redshift, \ariii/\oiii). We found that the acquisition rate for MPGs  reduced  by a few percent
 when using only one method, 92.3$\%$ for the $T_e$- method, 90.9$\%$ for the $R_{23}$-method,
 96.2$\%$ for the $N2$-method , 96.2$\%$ for the $O3N2$-, 85.8$\%$ for the $Ne3O2$-method
 and 88.7$\%$ for the $Ar3O3$- method(See Table 1). All the oxygen abundance determination methods
based on these strong line ratios are reliable to a certain degree.
  In any case, it  is an essential parameter for redshift to identify MPGs because  it
 is vital  to make the accurate redshift correction when deriving the the flux of the strong emission line.

 It is impressive that the acquisition rate for MPGs by $N2$-method and $Ne3O2$-method are
 comparable to it using all 9 variables. It might imply that both $N2$-method and $Ne3O2$-method
 are   monotonic, free of internal reddening correction, and therefore
 superior to other oxygen abundance determination methods.

 We  add (\ha/\hb) line ratio  to each method for shown of  the influence
of internal reddening correction on identifying MPGs.  This parameter is probed because the internal
 reddening correction is a fundamental step for determining \zoh(Shi et al.2005, 2006).
 We find that the acquisition rate for MPGs increases from 85.8$\%$ to 88.1$\%$ for the $Ne3O2$-method
 , 90.9$\%$ to 92.5$\%$ for the $R_{23}$-method and 88.7$\%$ to 90.9$\%$ for the $Ar3O3$- method
  when adding (\ha/\hb) to them(See Table 1).
 The acquisition rate for MPGs  for the rest three methods( $T_e$-, $N2$-, $O3N2$- method)
   do not change  when adding (\ha/\hb) to them, which can be explained  that the uncertainty
   of the  internal  reddening correction is comparable to the uncertainty of $T_e$-method and
   \nii$\lambda$6583/\ha$\lambda$6563, \oiii$\lambda$5007/\hb)/(\nii$\lambda$6583/\ha
   flux ratio is not sensitive the  internal  reddening correction.

\begin{table}[]
\caption{The acquisition rate for MPGs as a function of the different variables.}
\label{one}
\begin{tabular}{|c|c|c|c|c|c|c|}

\hline\hline
      & $T_e$ & $N2$ & $O3N2$&$R_{23}$&$Ne3O2$&$Ar3O3$\\
\hline
Z$^1$ & 0.923 & 0.962&  0.962 & 0.909&  0.858 & 0.887 \\
Z+ \ha/\hb$^2$&0.915   &  0.960 & 0.960  & 0.925&  0.881&  0.909 \\

\hline\hline
\end{tabular}
\\
1: The strong emission line ratios for each method plus redshift to identify MPGs.
\\
2: The strong emission line ratios for each method plus redshift and (\ha/\hb) to identify MPGs.
\\
\end{table}

To show the changes in performance caused by the MPG threshold,  we plot in Fig.5 the  confusion matrices
 when the MPG threshold is fixed to \zoh=8.39(corresponding to 0.5Z$_{\sun}$)
 using all 9 variables:  redshift, $Ne3O2$,  \oiii$\lambda\lambda$4959,5007/\oiii$\lambda$4363, $N2$/\ha, 
 \oiii/\hb, \sii, \oii/\hb, \ha/\hb and \ariii/\oiii. There are 8671 MPGs when the
MPG threshold is fixed to \zoh=8.39 and we select  10,000  MRGs galaxies for \zoh=8.39 randomly
  to avoid systematic error caused by too many MRGs compared to MPGs.
It is shown that the MPGs acquisition rate decreases a few percent compared to the
 MPG threshold of \zoh=8.0. This may imply that \zoh=8.39 is a less suitable MPG threshold
 than \zoh=8.00.

\begin{figure*}
   \centering
   \includegraphics[scale=0.9,angle=0]{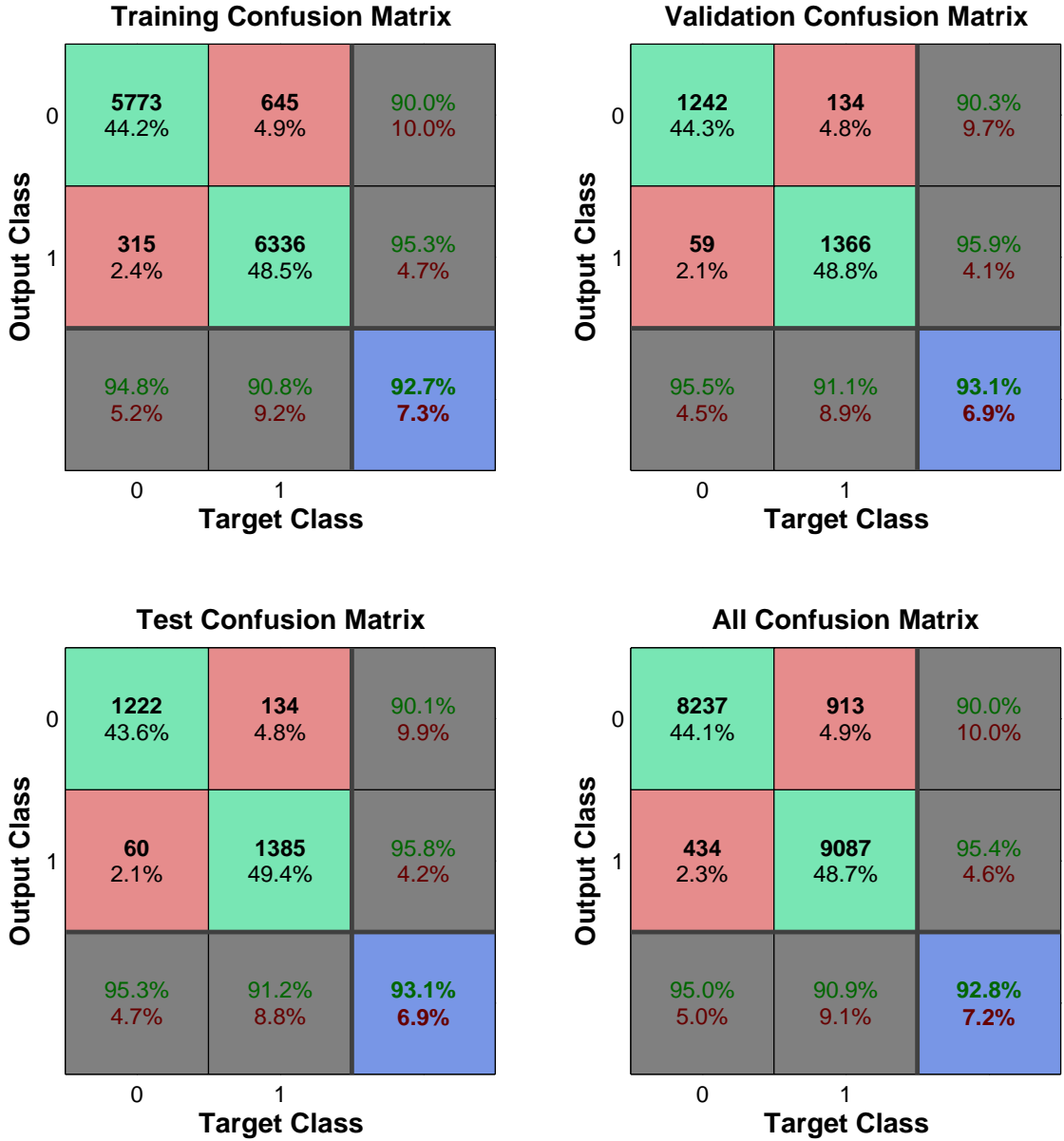}
   \caption{ The confusion matrices for training, testing, and validation,
    and the three kinds of data combined for the MPG threshold to \zoh=8.39 using all 9 variables.}
    \label{Fig8}%
    \end{figure*}

\section{Conclusions}

We have presented a promising new approach to select MPGs
from spectral catalogs. It involves the application of an artificial neural network with a
2-layer  feed-forward   architecture.
The input variables are spectral measurements, i.e., redshift and the most observably strong emission line ratios.

In the target selection, we have achieved an MPG acquisition rate
of 96 percent and 92 percent for an MPG threshold of \zoh=8.00 and \zoh=8.39,
respectively from $\sim$80,000 star forming galaxies.
{\bf The oxygen abundance of a galaxy in the MPG sample  have a 96 percent chance to be  lower than \zoh=8.00
 for an MPG threshold of \zoh=8.00.}
 
All the oxygen abundance determination methods
based on these strong line ratios are reliable to a certain degree, such as
the $T_e$-, $R_{23}$-,  $N2$- , $O3N2$-, $Ne3O2$- method,  and so on.
The acquisition rate for MPGs by $N2$-method and $O3N2$-method are
 comparable to it using all 9 variables. It shows  serious potential
 to search new MPGs candidate with a single emission line ratio,
 such as \nii$\lambda$6583/\ha$\lambda$6563.

This new statistical methods
developed in the context of the SDSS project can easily be
extended to any other analysis requiring MPG selection
when the physical property of the target can be quantitative.

Finally, we note that, aside from  its relative simplicity and robustness,
 the ANN classification method that we have presented here can be extended and
improved in a number of ways, {\bf such as increase of  neuron number,
 adoption of three-layer network, or making the multi category classification.
One have to be cautioned that both the classification accuracy and run-time may change dramatically
 in these processes. }

\begin{acknowledgements}

This work was funded by the National Natural Science Foundation of
China (NSFC) (Grant Nos.~11203001, 11202003 and 10873012), the National Basic Research Program
 of China (973 Program) (Grant No.~2007CB815404), and the Chinese
Universities Scientific Fund (CUSF).

Funding for the Sloan Digital Sky Survey (SDSS) has been provided
by the Alfred P. Sloan Foundation, the Participating Institutions, the
National Aeronautics and Space Administration, the National Science
Foundation, the U.S. Department of Energy, the Japanese Monbukagakusho,
and the Max Planck Society.
\end{acknowledgements}

\end{document}